\documentclass[a4paper]{article}

\usepackage{INTERSPEECH2021}
\DeclareMathOperator*{\argmax}{arg\,max}

\usepackage{algorithm}
\usepackage{algorithmic, multirow, hhline, graphicx, bbm}

\newlength{\bibitemsep}
\newlength{\bibparskip}
\let\oldthebibliography\thebibliography
\renewcommand\thebibliography[1]{%
  \oldthebibliography{#1}%
  \setlength{\parskip}{\bibitemsep}%
  \setlength{\itemsep}{\bibparskip}%
}

\title{Minimum Word Error Rate Training with Language Model Fusion for End-to-End Speech Recognition}
\name{Zhong Meng, Yu Wu, Naoyuki Kanda, Liang Lu, Xie Chen, Guoli Ye, \\ Eric Sun, Jinyu Li, Yifan Gong}
\address{Microsoft Corporation, Redmond, WA, USA}
\email{\{zhme,yuwu1,nakanda,lial,xieche,guoye,ersun,jinyli,ygong\}@microsoft.com}

\begin{document}

\maketitle
\begin{abstract}

Integrating external language models (LMs) into end-to-end (E2E) models remains a challenging task for domain-adaptive speech recognition. Recently, internal language model estimation (ILME)-based LM fusion has shown significant word error rate (WER) reduction from Shallow Fusion by subtracting a weighted internal LM score from an interpolation of E2E model and external LM scores during beam search. However, on different test sets, the optimal LM interpolation weights vary over a wide range and have to be tuned extensively on well-matched validation sets. In this work, we perform LM fusion in the minimum WER (MWER) training of an E2E model to obviate the need for LM weights tuning during inference. Besides MWER training with Shallow Fusion (MWER-SF), we propose a novel MWER training with ILME (MWER-ILME) where the ILME-based fusion is conducted to generate N-best hypotheses and their posteriors. Additional gradient is induced when internal LM is engaged in MWER-ILME loss computation.   
During inference, LM weights pre-determined in MWER training enable robust LM integrations on test sets from different domains. Experimented with 30K-hour trained transformer transducers, MWER-ILME achieves on average 8.8\% and 5.8\% relative WER reductions from MWER and MWER-SF training, respectively, on 6 different test sets. 

\end{abstract}
\noindent\textbf{Index Terms}: speech recognition, language model, transformer-transducer, end-to-end models

\section{Introduction}
End-to-end (E2E) models have achieved state-of-the-art performance for automatic speech recognition (ASR) \cite{chiu2018state, sainath2020streaming, li2020developing}. 
The most popular E2E models include connectionist temporal classification (CTC) \cite{graves2006connectionist, hannun2014deep, Li18CTCnoOOV}, recurrent neural network transducer (RNN-T) \cite{graves2012sequence, he2019streaming, Li2019RNNT} and attention-based encoder-decoder (AED) models \cite{chorowski2015attention, chan2016listen, karita2019comparative, Li2020Comparison}. E2E models are commonly trained to maximize the log posteriors of token sequences given speech sequences while the ASR performance is measured by the word error rate (WER).
Therefore, a minimum WER (MWER) criterion was proposed to train CTC \cite{graves2014towards}, AED \cite{prabhavalkar2018minimum}, RNN-T \cite{weng2019minimum, guo2020efficient} and hybrid autoregressive transducer (HAT) \cite{lu2020minimum} models, leading to improved ASR performance. 

However, an E2E model is more susceptible to domain shift \cite{belladaptation} from training to testing than a hybrid system \cite{DNN4ASR-hinton2012}.
Numerous methods have been proposed to adapt ASR models, such as regularization methods \cite{kld_yu, meng2019asa,l2_liao, meng2020lvector}, teacher-student learning \cite{li2017large, meng2018adversarial, manohar2018teacher, meng2019conditional}, transformation methods \cite{lhn, tan2015cluster}, and adversarial learning \cite{grl_shinohara, meng2018speaker, grl_serdyuk, dsn_meng}. However, all these methods require audio as the adaptation data when applied to E2E models \cite{ochiai2018speaker, meng2019speaker, meng2019domain}.
A promising solution is to integrate an external language model (LM) into the E2E model during inference or during MWER training \cite{peyser2020improving, weng2019minimum, kanda2017minimum}.
With no clear separation of acoustic and language models in an E2E model, LM fusion remains to be a challenging task.

Shallow Fusion \cite{hannun2014deep, gulcehre2015on, chorowski2016towards} is a simple yet effective LM fusion method which combines the probabilities of an E2E model and an LM through a log-linear interpolation at each step of beam search. 
The Density Ratio method \cite{mcdermott2019density, kanda2016maximum} improves Shallow Fusion by subtracting a source-domain LM score from the Shallow Fusion. 
Recently, inspired by HAT \cite{variani2020hybrid}, we proposed an internal LM estimation (ILME)-based Fusion \cite{meng2021ilme} which estimates the internal LM score of an E2E model and subtracts it from the Shallow Fusion score. 
ILME-based Fusion has shown consistent WER reductions from both Shallow Fusion and Density Ratio methods. Further, we proposed an internal LM training (ILMT) \cite{meng2021ilmt} that minimizes an internal LM loss in addition to the standard E2E loss. 
ILMT enables a more effective ILME-based Fusion during inference. 
However, with these methods, large WER reductions are only obtained when we carefully tune the LM interpolation weights on well-matched validation sets because the optimal LM weights often fluctuate dramatically on different test sets.

In this work, we perform LM fusion in the MWER training of E2E models to obviate the need for LM weights tuning. 
We first apply Shallow Fusion to generate the N-best hypotheses and compute their posteriors for MWER training (i.e., MWER-SF). 
Note that our MWER-SF differs from \cite{peyser2020improving} in that the E2E and LM scores are interpolated in the log domain and the combined scores are re-normalized over N-best hypotheses. 
Further, we propose a MWER training with ILME (MWER-ILME) in which the N-best hypotheses are generated by an ILME-based Fusion and their posteriors are computed by the probabilities of the E2E model, internal LM and external LM. The participation of internal LM in MWER-ILME loss computation induces additional gradient.
During inference, the LM fusion with LM weights preset in training are expected to achieve steady performance improvement on multiple different test sets.
Experimented with a 30 thousand (K)-hour trained transformer transducer, MWER-ILME achieves on average 8.8\% and 5.8\% relative WER reductions from MWER and MWER-SF, respectively, on a multi-domain test set, and the reductions are consistent over all subsets from different domains.

\section{Related Work}
\subsection{Minimum Word Error Rate Training}

Given a sequence of speech features $\mathbf{X}=\{\mathbf{x}_1,
\ldots, \mathbf{x}_T\}$ where $\mathbf{x}_t \in \mathbbm{R}^d$, an E2E model is typically trained to minimize the negative log posterior of the reference token sequence $\mathbf{Y^*}=\{y^*_1, \ldots, y^*_U\}$ where $y^*_u \in \mathcal{V}$ and $\mathcal{V}$ is the set of output tokens 
\begin{align}
   \mathcal{L}_\text{E2E}(\mathbf{X}, \mathbf{Y}^*) = -\log P(\mathbf{Y}^* | \mathbf{X};\theta^\text{S}_\text{E2E}), \label{eqn:e2e_loss}
\end{align}
where $\theta^\text{S}_\text{E2E}$ is the parameters of a source-domain E2E model.

With MWER training \cite{graves2014towards, prabhavalkar2018minimum, weng2019minimum}, the E2E model is further fine-tuned to directly minimize the expected number of word errors on the training corpus.
As a closer objective to the ASR performance metric, the MWER training is expected to outperform the standard E2E training. 
The MWER loss is commonly approximated as the expected number of word errors over the top N hypotheses $\{\mathbf{Y}^1,\ldots, \mathbf{Y}^N\}$ of $\mathbf{X}$ as follows
\begin{align}
    \mathcal{L}_\text{MWER}=&\sum_{n=1}^N \bar{P}(\mathbf{Y}^n | \mathbf{X};\theta^\text{S}_\text{E2E})R(\mathbf{Y}^n, \mathbf{Y}^*),
\end{align}
where $\bar{P}(\mathbf{Y}^n | \mathbf{X};\theta^\text{S}_\text{E2E}) = \frac{P(\mathbf{Y}^n | \mathbf{X};\theta^\text{S}_\text{E2E})}{\sum_{n=1}^N P(\mathbf{Y}^n | \mathbf{X};\theta^\text{S}_\text{E2E})} $ is the re-normalized posterior over the N-best hypotheses, and $R(\mathbf{Y}^n, \mathbf{Y}^*)$ is the number of word errors in a hypothesis $\mathbf{Y}^n$ compared to the reference $\mathbf{Y}^*$. 

\subsection{Shallow Fusion}
During inference, Shallow Fusion \cite{hannun2014deep, gulcehre2015on} searches for an optimal hypothesis $\hat{\mathbf{Y}}$ that maximizes a log-linear interpolation between the E2E model and the external LM probabilities. 
\begin{align}
    \hspace{-4pt} \hat{\mathbf{Y}} = \argmax_{\mathbf{Y}} \left[\log P(\mathbf{Y}|\mathbf{X};\theta^\text{S}_\text{E2E}) + \lambda_\text{T} \log P(\mathbf{Y};\theta^\text{T}_\text{LM}) \right],
\end{align}
where $P(\mathbf{Y};\theta^\text{T}_\text{LM})$ is the probability of an external LM trained with target-domain text, and $\lambda_\text{T}$ is the interpolation weight for the external LM.

\subsection{Internal LM Estimation-based Fusion}

An E2E model implicitly learns an internal LM 
$P(\mathbf{Y}^n;\theta^\text{S}_\text{E2E})$
that characterizes the distribution of source-domain training text. The internal LM probability $P(\mathbf{Y}; \theta^\text{S}_\text{E2E})$ can be estimated as the output of an E2E model by removing the contribution of the acoustic encoder (with parameters $\theta^\text{S}_\text{enc}$) \cite{variani2020hybrid, meng2021ilme}.

With the internal LM, we are able to compute the hypothesis posterior predicted by a target-domain E2E model $\theta^\text{T}_\text{E2E}$ with Bayes' Theorem as follows \cite{meng2021ilme}: 
\begin{align}
    \hspace{-2pt} P(\mathbf{Y} | \mathbf{X};\theta^\text{T}_\text{E2E})=&P(\mathbf{Y} | \mathbf{X};\theta^\text{S}_\text{E2E})\frac{P(\mathbf{Y};\theta^\text{T}_\text{LM})^{\lambda_\text{T}}}{P(\mathbf{Y};\theta^\text{S}_\text{E2E})^{\lambda_\text{S}}}g(\mathbf{X}), \label{eqn:bayes}
\end{align}
where $g(\mathbf{X}) = P(\mathbf{X};\theta^\text{S}_\text{E2E}) / P(\mathbf{X};\theta^\text{T}_\text{E2E}) = P_\text{S}(\mathbf{X}) / P_\text{T}(\mathbf{X})$ is the ratio of acoustic priors shared by all hypotheses.
ILME-based Fusion \cite{meng2021ilme} searches for the optimal hypothesis $\hat{\mathbf{Y}}$ with highest target-domain posterior $P(\mathbf{Y} | \mathbf{X};\theta^\text{T}_\text{E2E})$ as follows: 
\begin{align}
    \hat{\mathbf{Y}} = \argmax_{\mathbf{Y}} & \left[\log P(\mathbf{Y}|\mathbf{X}; \theta^\text{S}_\text{E2E}) + \lambda_\text{T} \log P(\mathbf{Y}; \theta^\text{T}_\text{LM}) \right. \nonumber \\
                     & \left. \quad - \lambda_\text{S} \log P(\mathbf{Y}; \theta^\text{S}_\text{E2E}) \right], \label{eqn:ilme}
\end{align}
where $\lambda_\text{S}$ is the interpolation weight for the internal LM.

In \cite{meng2021ilmt}, ILMT was proposed to encourage the acoustically-conditioned LM of an E2E model with parameters $\theta_\text{E2E}^\text{S} \setminus \theta_\text{enc}^\text{S}$ (e.g., the predictor and joint network of a transducer \cite{chen2020developing}, the decoder of AED \cite{meng2019character, gaur2019acoustic}) to also behave like a standalone LM, and thus facilitates a better external LM integration.
In ILMT training, an internal LM loss below is jointly minimized with the E2E loss $\mathcal{L}_\text{E2E}(\mathbf{X}, \mathbf{Y}^n)$ by updating the E2E model
\begin{align}
 \mathcal{L}_{\text{ILM}}(\mathbf{Y}^n) & = - \log P(\mathbf{Y}; \theta^\text{S}_\text{E2E}) \nonumber \\
 & = - \sum^{U + 1}_{u=1}\log P(y_u|\mathbf{Y}_{0:u-1};\theta_\text{E2E}^\text{S} \setminus \theta_\text{enc}^\text{S}), \label{eqn:ilm_loss}
\end{align}

\section{Minimum Word Error Rate Training with LM Fusion}
LM Fusion has remarkably improved the ASR performance of an E2E model with optimal LM weights carefully tuned on a validation set. However, the optimal LM interpolation weights fluctuate dramatically on different test sets. It may even happen that the optimal LM weights on one test set degrade the E2E model performance on the other set. Therefore, for each test set, the search for optimal LM weights relies on extensive tuning on a well-matched validation set, taking huge computational resources and human efforts. In many scenarios, such compatible validation sets are not even available. To obviate the need for LM weight tuning, we perform LM fusion in MWER training of E2E models.

\subsection{MWER Training with Shallow Fusion}
In \cite{peyser2020improving}, Shallow Fusion is performed during MWER training to generate N-best hypotheses $\{\mathbf{Y}^1,\ldots, \mathbf{Y}^N\}$ of $\mathbf{X}$. The posterior of each hypothesis is computed by a linear combination of E2E model and LM probabilities. 
Different from \cite{peyser2020improving}, the hypothesis posterior in this work is obtained by interpolating the log-probabilities of the E2E model and the external LM to better match the Shallow Fusion inference, and the word errors of each hypothesis is weighted by a re-normalized probability. Therefore, the MWER-SF loss is formulated as
\begin{align}
    \mathcal{L}^\text{SF}_\text{MWER}=&\sum_{n=1}^N \bar{P}(\mathbf{Y}^n | \mathbf{X};\theta^\text{S}_\text{E2E}, \theta^\text{T}_\text{LM})R(\mathbf{Y}^n, \mathbf{Y}^*),
\end{align}
where $\bar{P}(\mathbf{Y}^n | \mathbf{X}; \theta^\text{T}_\text{E2E}, \theta^\text{T}_\text{LM})$ is the re-normalized Shallow Fusion probability over N-best hypotheses below
\begin{align}
    \bar{P}(\mathbf{Y}^n | \mathbf{X}; \theta^\text{S}_\text{E2E}, \theta^\text{T}_\text{LM}) & = \frac{P(\mathbf{Y}^n | \mathbf{X};\theta^\text{S}_\text{E2E})P(\mathbf{Y}^n | \theta^\text{T}_\text{LM})^{\lambda_\text{T}}}{\sum_{i=1}^N P(\mathbf{Y}^{i} | \mathbf{X};\theta^\text{S}_\text{E2E})P(\mathbf{Y}^i | \theta^\text{T}_\text{LM})^{\lambda_\text{T}}}. \label{eqn:mwer_sf_posterior_norm}
\end{align}
During inference, the pre-defined LM weight in MWER-SF training is used for Shallow Fusion. MWER-SF adapts the E2E model to a fixed external LM and its interpolation weight, and mitigates the importance of the LM weight tuning.

\subsection{MWER Training with Internal LM Estimation}

We propose an MWER training of E2E models with ILME (MWER-ILME) by performing ILME-based Fusion in MWER training. MWER-ILME differs from standard MWER training in that: 1) we apply ILME-based Fusion to generate the N-best hypotheses of training utterances; 2) we compute hypothesis posteriors using probabilities of the E2E model, internal LM and external LM.

Therefore, an MWER-ILME loss function is defined as: 
\begin{align}
    \mathcal{L}^\text{ILME}_\text{MWER}&=\sum_{n=1}^N \bar{P}(\mathbf{Y}^n | \mathbf{X};\theta^\text{T}_\text{E2E})R(\mathbf{Y}^n, \mathbf{Y}^*).  \label{eqn:mwer_ilme_loss} 
\end{align}
Here, $\{\mathbf{Y}^1,\ldots, \mathbf{Y}^N\}$ are the N-best hypotheses of $\mathbf{X}$ generated by ILME-based Fusion performed with a source-domain E2E model, an internal LM and a fixed external LM. 
$\bar{P}(\mathbf{Y}^n | \mathbf{X}; \theta^\text{T}_\text{E2E})$ is the re-normalized posterior over N-best hypotheses below:
\begin{align}
    \bar{P}(\mathbf{Y}^n | \mathbf{X}; \theta^\text{T}_\text{E2E}) & = \frac{P(\mathbf{Y}^n | \mathbf{X};\theta^\text{T}_\text{E2E})}{\sum_{i=1}^N P(\mathbf{Y}^{i} | \mathbf{X};\theta^\text{T}_\text{E2E})}. \label{eqn:posterior_norm}
\end{align}
The posteriors of N-best hypotheses are computed by Eq. \eqref{eqn:bayes} as 
\begin{align}
    \hspace{-2pt} P(\mathbf{Y}^n | \mathbf{X};\theta^\text{T}_\text{E2E})=&P(\mathbf{Y}^n | \mathbf{X};\theta^\text{S}_\text{E2E})\frac{P(\mathbf{Y}^n;\theta^\text{T}_\text{LM})^{\lambda_\text{T}}}{P(\mathbf{Y}^n;\theta^\text{S}_\text{E2E})^{\lambda_\text{S}}}g(\mathbf{X}). \label{eqn:bayes_n}
\end{align}
MWER-ILME training is essentially minimizing the expected number of word errors over the N-best hypotheses generated by a target-domain E2E model $\theta^\text{T}_\text{E2E}$. 

With Eqs. \eqref{eqn:mwer_ilme_loss} to \eqref{eqn:bayes_n}, we express the MWER-ILME loss by $P(\mathbf{Y}^n | \mathbf{X};\theta^\text{S}_\text{E2E})$, $P(\mathbf{Y}^n;\theta^\text{T}_\text{LM})$ and $P(\mathbf{Y}^n;\theta^\text{S}_\text{E2E})$.
Note that, during MWER-ILME training, the only learnable parameters are $\theta^S_{\text{E2E}}$, and the inclusion of internal LM $P(\mathbf{Y}^n;\theta^\text{S}_\text{E2E})$ in loss function results in additional gradient compared to MWER training.

The derivative of MWER-ILME loss with respect to $\theta^\text{S}_{\text{E2E}}$ is
\begin{align}
\frac{\partial\mathcal{L}^\text{ILME}_{\text{MWER}}}{\partial \theta^\text{S}_\text{E2E}}
=\sum_{n=1}^N \frac{\partial\mathcal{L}^\text{ILME}_{\text{MWER}}}{\partial \log P(\mathbf{Y}^n | \mathbf{X};\theta^\text{T}_\text{E2E})} 
\frac{\partial \log P(\mathbf{Y}^n | \mathbf{X};\theta^\text{T}_\text{E2E})}{\partial \theta^S_\text{E2E}}. \label{eqn:mwer_ilme_grad}
\end{align}
In Eq. \eqref{eqn:mwer_ilme_grad}, the gradient of MWER-ILME loss with respect to the target-domain hypothesis posterior is computed as
\begin{align}
\frac{\partial\mathcal{L}^\text{ILME}_\text{MWER}}{\partial \log P(\mathbf{Y}^n | \mathbf{X};\theta^\text{T}_\text{E2E})} 
&= \bar{P}(\mathbf{Y}^n | \mathbf{X};\theta^\text{T}_\text{E2E})\left[R(\mathbf{Y}^n, \mathbf{Y}^*) - \bar{R}\right], \label{eqn:hyp_grad}
\end{align}
where $\bar{R}= \sum_{n=1}^N \bar{P}(\mathbf{Y}^n | \mathbf{X};\theta^\text{T}_\text{E2E})R(\mathbf{Y}^n, \mathbf{Y}^*)$ is the expected number of word errors over the N-best hypotheses. 
Given Eq. \eqref{eqn:bayes}, 
the gradient of the target-domain hypothesis posterior with respect to $\theta^S_\text{E2E}$ in Eq. \eqref{eqn:mwer_ilme_grad} can be expressed as
\begin{align}
& \frac{\partial \log P(\mathbf{Y}^n | \mathbf{X};\theta^\text{T}_\text{E2E})}{\partial \theta^\text{S}_\text{E2E}} \nonumber \\
& =\frac{\partial \log P(\mathbf{Y}^n | \mathbf{X};\theta^\text{S}_\text{E2E})}{\partial \theta^\text{S}_\text{E2E}} - \lambda_\text{S} \frac{\partial \log P(\mathbf{Y}^n ;\theta^\text{S}_\text{E2E})}{\partial \theta^\text{S}_\text{E2E}} \nonumber \\
& = - \frac{\partial \mathcal{L}_\text{E2E}(\mathbf{X}, \mathbf{Y}^n)}{\partial \theta^\text{S}_\text{E2E}}  + \lambda_\text{S} \frac{\partial \mathcal{L}_\text{ILM}(\mathbf{Y}^n)}{\partial \left(\theta^\text{S}_\text{E2E} \setminus \theta^\text{S}_\text{enc}\right)}, \label{eqn:params_grad}
\end{align}
where $\mathcal{L}_\text{E2E}(\mathbf{X}, \mathbf{Y}^n)$ and  $\mathcal{L}_\text{ILM}(\mathbf{Y}^n)$ are E2E and ILM losses in Eq. \eqref{eqn:e2e_loss} and Eq. \eqref{eqn:ilm_loss}, respectively. $\mathcal{L}_\text{E2E}$ for AED and $\mathcal{L}_\text{ILM}$ are both cross-entropy losses with simple gradients. For a transducer model, the gradient of  $\mathcal{L}_\text{E2E}$ is derived in \cite{graves2012sequence}. Therefore, the gradient of MWER-ILME loss with respect to $\theta^S_\text{E2E}$ is obtained by substituting Eqs. \eqref{eqn:hyp_grad} and \eqref{eqn:params_grad} into Eq. \eqref{eqn:mwer_ilme_grad}.

Through MWER-ILME training, we adapt the E2E model towards a fixed external LM along with the internal LM weight $\lambda_\text{S}$ and external LM weight $\lambda_\text{T}$. 
Therefore, during evaluation, the E2E model adapted by a \emph{multi-domain} LM is expected to get robust WER reductions on multiple different test sets if ILME-based Fusion is performed with the same set of LM weights.

\section{Experiments}
In this work, we train streaming transformer transducer (T-T) models \cite{chen2020developing} to minimize transducer \cite{graves2012sequence}, MWER \cite{prabhavalkar2018minimum}, MWER-SF and MWER-ILME losses, and evaluate it on a multi-domain test set. A multi-domain external LM is used in LM fusion for both MWER-SF and MWER-ILME training and evaluation. 
We also investigate the generalizability of MWER-SF and MWER-ILME training by evaluating with an out-of-domain LM on an out-of-domain test set. 

In all experiments, beam search and N-best generation are both performed with a beam size of 5. 3999 word-piece units generated by byte-pair encoding \cite{sennrich2015neural} are used as the output token set $\mathcal{V}$ for both T-T and the external LM. A word-count weighted average of subset WERs is computed for each test set.


\subsection{Language Models Training}
We use the same external LMs as in \cite{meng2021ilme, meng2021ilmt}. 

\textbf{Multi-domain LM}: We first train a uni-directional long short-term memory (LSTM)
\cite{sak2014long, erdogan2016multi, meng2017deep} LM on 2 billion words of anonymized text corpus comprising conversational data such as talks, interviews, meeting transcripts, and short message dictation from Microsoft services. 
The multi-domain LM has 2 hidden layers with 2048 units for each layer and in total 58 million (M) parameters.

\textbf{LibriSpeech LM}: We train another LSTM-LM with the 9.4M-word transcript of the 960-hour training speech and the 813M-word text in LibriSpeech corpus \cite{panayotov2015librispeech}. The LibriSpeech LM has exactly the same architecture as the multi-domain LM.



\subsection{Transformer Transducer Models Training}
\label{sec:tt}

We train T-T models with 30 thousand (K) hours of anonymized and transcribed data as in \cite{meng2021ilme, meng2021ilmt} collected from Microsoft services,
including voice search, short message dictation, conversations, command and control recorded in various conditions.

The T-T encoder starts with four 2D convolution layers that sub-sample the input 
80-dimensional (dim) log Mel-filter bank 
features in time by a factor of 4 and project them to 512 dimensions. This is followed by a transformer with 18 layers, each layer has an 8-head attention sub-layer with relative positional encoding \cite{vaswani2017attention} and a 2048-dim fully-connected sub-layer. The encoder has a look-ahead of 360 ms on average. The T-T predictor is a 2-layer
transformer with each layer containing a 4-head attention sub-layer followed by a 1024-dim fully-connected sub-layer. 
The input to the predictor are 512-dim word-piece embedding vectors with positional encoding. 
The attention dimension is fixed at 512 for both encoder and predictor. The outputs of encoder and predictor are projected to 512-dim vectors.
Dropout \cite{srivastava2014dropout} with a probability of 0.1 is deployed in both the encoder and the predictor. The T-T model has 67 million (M) parameters. 

First, a T-T model is trained to minimize the transducer loss with the 30K-hour data.
The baseline T-T is then fine-tuned with the same data to minimize MWER, MWER-SF and MWER-ILME losses using Adam optimizer \cite{kingma2014adam} with a constant learning rate of $10^{-5}$. The multi-domain LM is used as the external LM for both MWER-SF and MWER-ILME training. In MWER-SF training, the external LM weight is set to 0.25. In MWER-ILME training, the external and internal LM weights are set to 0.25 and 0.05, respectively.

\subsection{Multi-Domain Evaluations}
\label{sec:multi_domain_eval}
We collect a multi-domain test set containing 6 subsets from domains covered by the multi-domain LM to evaluate T-Ts trained in Section \ref{sec:tt} as follows.

\textbf{Call:} 30K-word conversational speech collected from real online phone calls.

\textbf{Meeting:} 5K-word conversational speech collected from real meetings.

\textbf{Search:} 29K-word voice search speech collected from multiple microphone arrays.

\textbf{Keyboard:} 15K-word dictated speech collected from keyboard voice input.

\textbf{Email:} 9K-word dictated speech from email applications.

\textbf{Common:} the Common Voice \cite{ardila2019common} test set consisting of 125k-word spoken text from blog posts, old books, movies, etc. \\
Except for Common, the other five anonymized subsets are all collected from Microsoft services. 

The baseline and MWER-trained T-Ts are evaluated on the muliti-domain test set \emph{without} using external LM. As shown in Table \ref{table:multi_domain_wer}, the baseline T-T achieves 12.00\% WER on average over 6 subsets. MWER training leads to 4.8\% relative WER reduction from the baseline T-T.
\begin{table}
\centering
\setlength{\tabcolsep}{4.0pt}
\begin{tabular}[c]{c||c|c|c|c||c}
	\hline
	\hline
	\multirow{2}{*}{\begin{tabular}{@{}c@{}} Test \\ Subset \end{tabular}}  
	& \multirow{2}{*}{\begin{tabular}{@{}c@{}} T-T \end{tabular}} 
	& \multirow{2}{*}{\begin{tabular}{@{}c@{}} MWER \end{tabular}} 
	& \multirow{2}{*}{\begin{tabular}{@{}c@{}} MWER- \\ SF \end{tabular}} 
	& \multirow{2}{*}{\begin{tabular}{@{}c@{}} MWER- \\ ILME \end{tabular}} 
	& \multirow{2}{*}{\begin{tabular}{@{}c@{}} WERR \end{tabular}} \\
	\hhline{~~~~~~}
	& & & & & \\
	\hline
    Call & 8.70 & 8.37 & 7.70 & \textbf{7.58} & 9.4 \\
	\hline
    Meeting & 16.34 & 16.07 & 17.54 & \textbf{15.87} & 1.2 \\
	\hline
    Search & 12.56 & 12.35 & 12.13 & \textbf{11.73} & 5.0 \\
	\hline
    Keyboard & 7.95 & 7.56 & 7.60 & \textbf{7.37} & 2.5 \\
	\hline
    Email & 19.16 & 17.67 & 16.94 & \textbf{16.31} & 7.7 \\
	\hline
    Common & 12.43 & 11.75 & 11.34 & \textbf{10.51} & 10.6 \\
	\hline
    Avg. & 12.00 & 11.43 & 11.07 & \textbf{10.43} & 8.8 \\
	\hline
	\hline
	\end{tabular}
	\caption{WERs (\%) of T-T models trained with 4 different objectives on a \textbf{multi-domain} test set including 6 subsets. WERR (\%) is the relative WER reduction of MWER-ILME from MWER.}
\label{table:multi_domain_wer}
\vspace{-15 pt}
\end{table}
In Table \ref{table:multi_domain_oracle_wer}, we evaluate the MWER-trained T-T with Shallow Fusion and ILME-based Fusion. We first tune the LM weights on the Call subset and apply the optimal weights for all 6 subsets (i.e., Oracle Call). Although both LM fusions achieve significant WER reductions on Call and Search, the performance degrades severely on Meeting, Keyboard and Common. 
This indicates that the optimal LM weights fluctuate dramatically on different test sets.
Further, we conduct LM fusions on each subset with the optimal LM weights tuned on itself (i.e., Oracle All). Both LM fusions result in consistent WER reductions on all subsets, with ILME-based Fusion performing better on each subset.

\begin{table}
\centering
\setlength{\tabcolsep}{5.3pt}
\begin{tabular}[c]{c||c|c|c|c||c}
	\hline
	\hline
	\multirow{2}{*}{\begin{tabular}{@{}c@{}} Test \\ Subset \end{tabular}}
	& \multirow{2}{*}{\begin{tabular}{@{}c@{}} No LM \end{tabular}}
	& \multicolumn{2}{c|}{Oracle Call}
	& \multicolumn{2}{c}{Oracle All} \\
	\hhline{~~----}
	& & SF & ILME & SF & ILME \\
	\hline
    Call & 8.37 &	7.74 & 7.33 & 7.74 & 7.33 \\
	\hline
    Meeting & 16.07 & 18.99 & 17.98 & 15.97 & 15.78 \\
	\hline
    Search & 12.35 & 12.27 & 12.01 & 11.97 & 11.66 \\
	\hline
    Keyboard & 7.56 & 7.92 & 7.92 & 7.54 & 7.37 \\
	\hline
    Email & 17.67 & 17.85 & 16.55 & 16.74 & 16.19 \\
	\hline
    Common & 11.75 & 12.35 & 11.94 & 11.33 & 11.05 \\
	\hline
    Avg. & 11.43 & 11.78 & 11.37 & 11.00 & 10.69 \\
	\hline
	\hline
	\end{tabular}
	\caption{WERs (\%) of MWER trained T-T model on a \textbf{multi-domain} test set. ``Oracle Call'' uses optimal LM weights tuned on Call subset for LM fusion on all 6 subsets. ``Oracle All'' uses optimal LM  weights tuned on each subset for LM fusion. }
\label{table:multi_domain_oracle_wer}
\vspace{-20 pt}
\end{table}

We then evaluate the MWER-SF trained T-T with Shallow Fusion by using the same multi-domain LM and LM weight (0.25) as in training. 
In Table \ref{table:multi_domain_wer}, although MWER-SF improves MWER training by 3.1\% relatively in terms of averaged WER, it degrades WERs on Meeting and Keyboard subsets by 9.1\% and 5.3\% relatively, respectively.
Finally, we evaluate the MWER-ILME trained T-T with ILME-based Fusion by using the same multi-domain LM and external/internal LM weights (0.25/0.05) as in training. Among all 4 training objectives, MWER-ILME performs the best on every subset of the multi-domain test set, achieving on average 8.8\% and 5.8\% relative WER reductions from the MWER and MWER-SF training, respectively.  MWER-ILME consistently improves MWER training by 1.2\% to 10.6\% relatively on 6 subsets with no performance degradation. 
With MWER-ILME training, ILME-based Fusion is able to achieve significant and robust relative WER reductions on test sets from multiple domains by applying the same multi-domain LM and LM weights during training. 

Note that MWER-ILME in Table \ref{table:multi_domain_wer} even outperforms the MWER-trained T-T with ILME-based Fusion and ``Oracle All''
in Table \ref{table:multi_domain_oracle_wer} by 2.4\% relatively on average. This shows that, during MWER-ILME, the adaptation of T-T towards external LM also facilitates a more effective fusion between these two models. The need for LM weight tuning is thus entirely eliminated. 



\subsection{Out-of-Domain Evaluations}
We further investigate the generalizability of MWER-SF and MWER-ILME trained T-Ts by integrating out-of-domain LMs into them, and performing the evaluation on an out-of-domain test set. 
We choose LiriSpeech LM and the LibriSpeech test set for out-of-domain evaluation since neither the training data of T-T nor that of the multi-domain LM contains LibriSpeech data. The LibriSpeech test set consists of 4 subsets, ``dev-clean'', ``test-clean'', ``dev-other'', ``test-other'', including 54K, 53K, 51K and 52K words, respectively \cite{panayotov2015librispeech}.

The baseline and MWER-trained T-Ts are evaluated \emph{without} external LM. 
The MWER-SF and MWER-ILME trained T-Ts are evaluated with Shallow Fusion and ILME-based Fusion, respectively. During evaluation, the LM fusions are performed with LibriSpeech LM and the same LM weights as in training. In Table \ref{table:librispeech_wer}, the MWER trained T-T achieves on average 9.94\% WER on LibriSpeech test set, with a 2.6\% relative WER reduction from baseline T-T. MWER-SF outperforms MWER training by 20.42\% relatively on average, and the improvement is consistent over all subsets. Among all 4 methods, MWER-ILME performs the best on every subset, achieving an on average 21.7\% relative WER reduction from MWER training. 

The results show that, even performed with a multi-domain LM, neither MWER-SF nor MWER-ILME training biases the T-T entirely towards the domains covered by the external LM. Instead, they equip T-Ts with the generalizability to be integrated with an out-of-domain LM and achieve a significant WER reduction on an out-of-domain test set.

\begin{table}
\centering
\setlength{\tabcolsep}{4.5pt}
\begin{tabular}[c]{c||c|c|c|c||c}
	\hline
	\hline
	\multirow{2}{*}{\begin{tabular}{@{}c@{}} Test Set \end{tabular}}  
	& \multirow{2}{*}{\begin{tabular}{@{}c@{}} T-T \end{tabular}} 
	& \multirow{2}{*}{\begin{tabular}{@{}c@{}} MWER \end{tabular}} 
	& \multirow{2}{*}{\begin{tabular}{@{}c@{}} MWER- \\ SF \end{tabular}} 
	& \multirow{2}{*}{\begin{tabular}{@{}c@{}} MWER- \\ ILME \end{tabular}} 
	& \multirow{2}{*}{\begin{tabular}{@{}c@{}} WERR \end{tabular}} \\
	\hhline{~~~~~~}
	& & & & & \\
	\hline
    dev-clean & 6.11 & 6.03 & 4.46 & \textbf{4.42} & 26.7 \\
	\hline
    dev-other & 14.34 & 13.87 & 11.29 & \textbf{11.09} & 20.0 \\
	\hline
    test-clean & 6.18 & 5.98 & 4.46 & \textbf{4.42} & 26.1 \\
	\hline
    test-other & 14.51 & 14.15 & 11.67 & \textbf{11.43} & 19.2 \\
	\hline
    Avg. & 10.21 & 9.94 & 7.91 & \textbf{7.78} & 21.7 \\
	\hline
	\hline
	\end{tabular}
	\caption{WERs (\%) of T-T models trained with 4 different objectives on \textbf{out-of-domain} LibriSpeech dev and test sets. WERR (\%) is the relative WER reduction of MWER-ILME from MWER.}
\label{table:librispeech_wer}
\vspace{-20 pt}
\end{table}

\section{Conclusion}
In this work, we perform the LM fusion in the MWER training of a T-T model to eliminate the need for LM weight tuning during inference. In addition to Shallow Fusion, we propose to apply ILME-based Fusion for the N-best generation and the hypothesis score computation in the MWER training. With LM weights preset in MWER-ILME training, ILME-based fusion significantly and consistently outperforms MWER and MWER-SF trained models on multiple different test sets, achieving on average 8.8\% and 5.8\% relative WER reductions, respectively.
MWER-ILME even outperforms MWER-trained model with ILME-based Fusion and \emph{oracle} LM weights.
With great generalizability, both MWER-SF and MWER-ILME trained E2E models perform more than 20\% relatively better than MWER trained model on an out-of-domain test set when integrated with an out-of-domain LM.


\bibliographystyle{IEEEtran}

\bibliography{refs}


\end{document}